\documentclass[10pt]{article}
\usepackage{hyperref,graphicx,bchart}
\usepackage[affil-it]{authblk}
\usepackage{titlesec}

\usepackage[letterpaper,margin=1.25in]{geometry}

\newcommand{\bn}{\begin{enumerate}}
\newcommand{\en}{\end{enumerate}}
\newcommand{\bi}{\begin{itemize}}
\newcommand{\ei}{\end{itemize}}

\titleformat{\section}{\large\bfseries}{\thesection}{1em}{}

\title{\vspace*{-2cm} {\large \textbf{A Snapshot of Foundational Attitudes Toward Quantum Mechanics}}} 

\author[1]{Maximilian Schlosshauer\thanks{E-mail: \texttt{schlossh@up.edu}}}

\author[2]{Johannes Kofler}

\author[3]{Anton Zeilinger}

\affil[1]{Department of Physics, University of Portland, 5000~North
  Willamette Boulevard, Portland,~Oregon~97203,~USA}

\affil[2]{Max Planck Institute of Quantum Optics, Hans-Kopfermann-Stra{\ss}e~1, 85748~Garching,~Germany}

\affil[3]{Institute for Quantum Optics and Quantum Information, Austrian Academy of Sciences, Boltzmanngasse~3, 1090~Vienna, Austria; and
Vienna Center for Quantum Science and Technology, Department of Physics, University of Vienna, Boltzmanngasse~5, 1090~Vienna, Austria}

\date{}

\begin{document}

\maketitle

\vspace*{-.7cm}

\begin{abstract}
\noindent Foundational investigations in quantum mechanics, both experimental and theoretical, gave birth to the field of quantum information science. Nevertheless, the foundations of quantum mechanics themselves remain hotly debated in the scientific community, and no consensus on essential questions has been reached. Here, we present the results of a poll carried out among 33 participants of a conference on the foundations of quantum mechanics. The participants completed a questionnaire containing 16 multiple-choice questions probing opinions on quantum-foundational issues. Participants included physicists, philosophers, and mathematicians. We describe our findings, identify commonly held views, and determine strong, medium, and weak correlations between the answers. Our study provides a unique snapshot of current views in the field of quantum foundations, as well as an analysis of the relationships between these views.
\end{abstract}

\section{Why this poll?}

In August 1997, Max Tegmark polled 48 participants of the conference ``Fundamental Problems in Quantum Theory,'' held at the University of Maryland, Baltimore County, about their favorite interpretation of quantum mechanics \cite{Tegmark:1998:qq}. By Tegmark's own admission, the survey was ``highly informal and unscientific,'' as ``several people voted more than once, many abstained, etc.'' While the Copenhagen interpretation gathered the most votes, the many-worlds interpretation turned out to come in second, prompting Tegmark to declare a ``rather striking shift in opinion compared to the old days when the Copenhagen interpretation reigned supreme.'' 

Today, debates about the foundations of quantum mechanics show no sign of abating. Indeed, they have only become more lively in the years since Tegmark's poll. Thus, we felt the time had come to take a new snapshot. A perfect photo opportunity had just presented itself: the conference ``Quantum Physics and the Nature of Reality,'' held in July 2011 at the International Academy Traunkirchen, Austria, and organized by one of us (A.Z.). A mix of physicists, philosophers, and mathematicians had gathered at a former monastery at the shore of Lake Traunsee in Austria (see Appendix~\ref{sec:list-conf-part} for a list of participants). We handed the conference participants a prepared questionnaire with 16 multiple-choice questions covering the main issues and open problems in the foundations of quantum mechanics. We permitted multiple answers per question to be checked, because in many cases the different answers were not, and could not be, mutually exclusive. 

Just as Tegmark's poll, our poll cannot claim to be representative of the communities at large. But, as a snapshot, it contains interesting---and in parts even surprising---information. A total of 33 people turned in their completed questionnaires; of those, 27 stated their main academic affiliation as physics, 5 as philosophy, and 3 as mathematics (here, too, multiple answers were allowed). While this is not a huge sample size, it is to our knowledge the most comprehensive poll of quantum-foundational views ever conducted.\footnote{An in-depth account of the different quantum-foundational views as told by some of the discipline's main protagonists, including respondents to this poll, can be found in \cite{Schlosshauer:2011:ee}. This book features interviews with seventeen physicists, philosophers, and mathematicians replying to a fixed set of questions about the central issues in quantum foundations.} Also, we were certainly aware of the fact that the multiple-choice format can sometimes obliterate the all-important nuances: two people may check the answer ``local realism is untenable,'' and yet mean completely different concepts by each word in this sentence. This, however, is a small price to pay for the ability to directly tally up the votes and to analyze correlations between answers.

In Sec.~\ref{sec:results}, we present the results of the poll. In Sec.~\ref{sec:corr}, we visualize and investigate correlations between different answers. In Sec.~\ref{sec:discussion}, we comment on our findings. In Appendix~\ref{sec:correlation-table}, we present a table showing correlations of different strengths between answers. In Appendix~\ref{sec:list-conf-part}, we list the names and affiliations of the conference participants.

\section{\label{sec:results}Results}

\subsubsection*{Question 1: What is your opinion about the randomness of individual quantum events (such as the decay of a radioactive atom)?}

\scalebox{0.85}{\begin{bchart}[step=10,max=100,unit=\%,width=0.8\textwidth]

\bcbar[value=,text={\emph{a.} The randomness is only apparent:}]{0}
\bcbar{9}\smallskip

\bcbar[value=,text={\emph{b.} There is a hidden determinism:}]{0}
\bcbar{0}\smallskip

\bcbar[value=,text={\emph{c.} The randomness is irreducible:}]{0}
\bcbar{48}\smallskip

\bcbar[value=,text={\emph{d.} Randomness is a fundamental concept in nature:}]{0}
\bcbar{64}

\bcxlabel{percent of votes}
\end{bchart}}

\noindent Although we did not elaborate on the meaning of the word ``apparent'' in the provided answer, the distinction between the first and the second answer becomes clear when one contrasts the Everett interpretation with hidden-variables theories such as the de~Broglie--Bohm interpretation. In the Everett interpretation, randomness is an apparent effect relative to a branching observer, with the global wave function (containing all branches) evolving deterministically according to the Schr\"odinger equation. This does not match the notion of a ``hidden determinism,'' which is commonly associated with ideas such as hidden variables, deterministic mechanisms underlying the occurrence of objective quantum events, and (as in the de~Broglie--Bohm interpretation) the deterministic motion of physical constituents. In our poll, none of the participants favored the de~Broglie--Bohm interpretation, but there was a fair amount of support for the Everett interpretation (see Question~12). This may explain why no votes were cast in favor of a hidden determinism, while the idea of apparent randomness found several supporters.

\subsubsection*{Question 2: Do you believe that physical objects have their properties well defined prior to and independent of measurement?}

\scalebox{0.85}{\begin{bchart}[step=10,max=100,unit=\%,width=0.8\textwidth]

\bcbar[value=,text={\emph{a.} Yes, in all cases:}]{0}
\bcbar{3}\smallskip

\bcbar[value=,text={\emph{b.} Yes, in some cases:}]{0}
\bcbar{52}\smallskip

\bcbar[value=,text={\emph{c.} No:}]{0}
\bcbar{48}\smallskip

\bcbar[value=,text={\emph{d.} I'm undecided:}]{0}
\bcbar{9}

\bcxlabel{percent of votes}
\end{bchart}}

\noindent There were some write-ins in response to the question. One respondent asked, ``What is `well defined'?'' Another called it a ``bad question, because it presumes local realism and ignores entanglement.''

\subsubsection*{Question 3: Einstein's view of quantum mechanics}

\scalebox{0.85}{\begin{bchart}[step=10,max=100,unit=\%,width=0.8\textwidth]

\bcbar[value=,text={\emph{a.} Is correct:}]{0}
\bcbar{0}\smallskip

\bcbar[value=,text={\emph{b.} Is wrong:}]{0}
\bcbar{64}\smallskip

\bcbar[value=,text={\emph{c.} Will ultimately turn out to be correct:}]{0}
\bcbar{6}\smallskip

\bcbar[value=,text={\emph{d.} Will ultimately turn out to be wrong:}]{0}
\bcbar{12}\smallskip

\bcbar[value=,text={\emph{e.} We'll have to wait and see:}]{0}
\bcbar{12}

\bcxlabel{percent of votes}
\end{bchart}}

\noindent In wording our question, we deliberately did not specify what exactly we took Einstein's view of quantum mechanics to be. It is well known, in fact, that Einstein held a variety of views over his lifetime \cite{Fine:1996:sg}. The overarching themes we were after---and the themes most people, we believe, would associate with Einstein---are a subtle flavor of realism, as well as the possibility of a deeper description of nature beneath quantum mechanics. 

Interestingly, none of the respondents brought himself to declaring Einstein's view as correct, although two people suggested that Einstein would ultimately be vindicated. One respondent sounded a conciliatory note. ``Einstein's view is wrong. But he still thought more clearly than anyone else in his time. There is still much to learn from him.''

\subsubsection*{Question 4: Bohr's view of quantum mechanics}

\scalebox{0.85}{\begin{bchart}[step=10,max=100,unit=\%,width=0.8\textwidth]

\bcbar[value=,text={\emph{a.} Is correct:}]{0}
\bcbar{21}\smallskip

\bcbar[value=,text={\emph{b.} Is wrong:}]{0}
\bcbar{27}\smallskip

\bcbar[value=,text={\emph{c.} Will ultimately turn out to be correct:}]{0}
\bcbar{9}\smallskip

\bcbar[value=,text={\emph{d.} Will ultimately turn out to be wrong:}]{0}
\bcbar{3}\smallskip

\bcbar[value=,text={\emph{e.} We'll have to wait and see:}]{0}
\bcbar{30}

\bcxlabel{percent of votes}
\end{bchart}}

\noindent Just as with the previous question about Einstein's view of quantum mechanics, we did not elaborate on the specifics of what we meant by ``Bohr's view.'' Bohr, of course, has become associated with a variety of positions, and it is likely that in responding to the question, each participant had a slightly different set of ideas and slogans in mind. 

The question prompted some inspired write-ins. Bohr's view, one respondent wrote, ``is correct \emph{and} wrong: he had good insights, badly expressed.'' Someone else called it ``correct, but narrow-minded.'' And a third responded: ``I still can't figure out what I think of this guy. I certainly like many elements from the thinking of his disciples.''

\subsubsection*{Question 5: The measurement problem}

\scalebox{0.85}{\begin{bchart}[step=10,max=100,unit=\%,width=0.8\textwidth]

\bcbar[value=,text={\emph{a.} A pseudoproblem:}]{0}
\bcbar{27}\smallskip

\bcbar[value=,text={\emph{b.} Solved by decoherence:}]{0}
\bcbar{15}\smallskip

\bcbar[value=,text={\emph{c.} Solved/will be solved in another way:}]{0}
\bcbar{39}\smallskip

\bcbar[value=,text={\emph{d.} A severe difficulty threatening quantum mechanics:}]{0}
\bcbar{24}\smallskip

\bcbar[value=,text={\emph{e.} None of the above:}]{0}
\bcbar{27}

\bcxlabel{percent of votes}
\end{bchart}}

\noindent Only a fourth of the respondents regarded the measurement problem as a ``severe difficulty.'' This is a noteworthy result, for the measurement problem is often portrayed as the main stumbling block for quantum theory, as a poisoned thorn in the side of the theory that ought to be removed before any further progress can occur.

\subsubsection*{Question 6: What is the message of the observed violations of Bell's inequalities?}

\noindent \scalebox{0.85}{\begin{bchart}[step=10,max=100,unit=\%,width=0.8\textwidth]

\bcbar[value=,text={\emph{a.} Local realism is untenable:}]{0}
\bcbar{64}\smallskip

\bcbar[value=,text={\emph{b.} Action-at-a-distance in the physical world:}]{0}
\bcbar{12}\smallskip

\bcbar[value=,text={\emph{c.} Some notion of nonlocality:}]{0}
\bcbar{36}\smallskip

\bcbar[value=,text={\emph{d.} Unperformed measurements have no results:}]{0}
\bcbar{52}\smallskip

\bcbar[value=,text={\emph{e.} Let's not jump the gun---let's take the loopholes more seriously:}]{0}
\bcbar{6}

\bcxlabel{percent of votes}
\end{bchart}}

\noindent The Bell inequalities are a wonderful example of how we can have a rigorous theoretical result tested by numerous experiments, and yet disagree about the implications. The results of our poll clearly support this observation.

\subsubsection*{Question 7: What about quantum information?}

\scalebox{0.85}{\begin{bchart}[step=10,max=100,unit=\%,width=0.8\textwidth]

\bcbar[value=,text={\emph{a.} It's a breath of fresh air for quantum foundations:}]{0}
\bcbar{76}\smallskip

\bcbar[value=,text={\emph{b.} It's useful for applications but of no relevance to quantum foundations:}]{0}
\bcbar{6}\smallskip

\bcbar[value=,text={\emph{c.} It's neither useful nor fundamentally relevant:}]{0}
\bcbar{6}\smallskip

\bcbar[value=,text={\emph{d.} We'll need to wait and see:}]{0}
\bcbar{27}

\bcxlabel{percent of votes}
\end{bchart}}

\noindent Evidently, there is broad enthusiasm---or at least open-mindedness---about quantum information, with three in four respondents regarding quantum information as  ``a breath of fresh air for quantum foundations.'' Indeed, it it hard to deny the impact quantum information theory has had on the field of quantum foundations over the past decade. It has inspired new ways of thinking about quantum theory and has produced information-theoretic derivations (reconstructions) of the structure of the theory. On the practical side, the quantum-information boom has helped fund numerous foundational research projects. Last but not least, quantum information has given foundational pursuits new legitimacy. 

\subsubsection*{Question 8: When will we have a working and useful quantum computer?}

\scalebox{0.85}{\begin{bchart}[step=10,max=100,unit=\%,width=0.8\textwidth]

\bcbar[value=,text={\emph{a.} Within 10 years:}]{0}
\bcbar{9}\smallskip

\bcbar[value=,text={\emph{d.} In 10 to 25 years:}]{0}
\bcbar{42}\smallskip

\bcbar[value=,text={\emph{c.} In 25 to 50 years:}]{0}
\bcbar{30}\smallskip

\bcbar[value=,text={\emph{d.} In 50 to 100 years:}]{0}
\bcbar{0}\smallskip

\bcbar[value=,text={\emph{e.} Never:}]{0}
\bcbar{15}

\bcxlabel{percent of votes}
\end{bchart}}

\noindent Generally, there seemed to be a sense of cautious optimism. The majority of those who gave explicit time estimates thought we would see quantum computers within the next 10 to 25 years, and no one expected it would take longer than 50 years. There was one write-in: ``We have one now---conscious beings!''

\subsubsection*{Question 9: What interpretation of quantum states do you prefer?}

\scalebox{0.85}{\begin{bchart}[step=10,max=100,unit=\%,width=0.8\textwidth]

\bcbar[value=,text={\emph{a.} Epistemic/informational:}]{0}
\bcbar{27}\smallskip

\bcbar[value=,text={\emph{b.} Ontic:}]{0}
\bcbar{24}\smallskip

\bcbar[value=,text={\emph{c.} A mix of epistemic and ontic:}]{0}
\bcbar{33}\smallskip

\bcbar[value=,text={\emph{d.} Purely statistical (e.g., ensemble interpretation):}]{0}
\bcbar{3}\smallskip

\bcbar[value=,text={\emph{e.} Other:}]{0}
\bcbar{12}

\bcxlabel{percent of votes}
\end{bchart}}

\noindent This is a perfect example of a question where the options are not well defined. We did not give a definition of notions like epistemic, informational, and ontic. Nonetheless, the distribution of answers gives a good picture---of the present situation, anyway. In light of the fact that epistemic and ontic interpretations constitute such radically different viewpoints, it is interesting to observe a virtual draw between these two interpretations. Remarkably, supporters of the ensemble interpretation seemed to have all but disappeared.

\subsubsection*{Question 10: The observer}

\scalebox{0.85}{\begin{bchart}[step=10,max=100,unit=\%,width=0.8\textwidth]

\bcbar[value=,text={\emph{a.} Is a complex (quantum) system:}]{0}
\bcbar{39}\smallskip

\bcbar[value=,text={\emph{b.} Should play no fundamental role whatsoever:}]{0}
\bcbar{21}\smallskip

\bcbar[value=,text={\emph{c.} Plays a fundamental role in the application of the formalism but plays no distinguished physical role:}]{0}
\bcbar{55}\smallskip

\bcbar[value=,text={\emph{d.} Plays a distinguished physical role (e.g., wave-function collapse by consciousness):}]{0}
\bcbar{6}

\bcxlabel{percent of votes}
\end{bchart}}

\noindent It is remarkable that more than 60\% of respondents appear to believe that the observer is not a complex quantum system. Also, very few adhere to the notion that the observer plays a distinguished physical role (for example, through a consciousness-induced collapse of the wave function). Given the relatively strong (42\%) support for the Copenhagen interpretation (see Question~12), this finding shows that support of the Copenhagen interpretation does not necessarily imply a belief in a fundamental role for consciousness. (Popular accounts have sometimes suggested that the Copenhagen interpretation attributes such a role to consciousness. In our view, this is to misunderstand the Copenhagen interpretation.)

\subsubsection*{Question 11: Reconstructions of quantum theory}

\scalebox{0.85}{\begin{bchart}[step=10,max=100,unit=\%,width=0.8\textwidth]

\bcbar[value=,text={\emph{a.} Give useful insights and have superseded/will supersede the interpretation program:}]{0}
\bcbar{15}\smallskip

\bcbar[value=,text={\emph{b.} Give useful insights, but we still need interpretation:}]{0}
\bcbar{45}\smallskip

\bcbar[value=,text={\emph{c.} Cannot solve the problems of quantum foundations:}]{0}
\bcbar{30}\smallskip

\bcbar[value=,text={\emph{d.} Will lead to a new theory deeper than quantum mechanics:}]{0}
\bcbar{27}\smallskip

\bcbar[value=,text={\emph{e.} Don't know:}]{0}
\bcbar{12}

\bcxlabel{percent of votes}
\end{bchart}}

\noindent It is remarkable that a great majority attributes a useful role to reconstruction of quantum theory. This is even more remarkable given that the research program of finding and developing a reconstruction of quantum theory is a rather young one.

\subsubsection*{Question 12: What is your favorite interpretation of quantum mechanics?}

\scalebox{0.85}{\begin{bchart}[step=10,max=100,unit=\%,width=0.8\textwidth]

\bcbar[value=,text={\emph{a.} Consistent histories:}]{0}
\bcbar{0}\smallskip

\bcbar[value=,text={\emph{b.} Copenhagen:}]{0}
\bcbar{42}\smallskip

\bcbar[value=,text={\emph{c.} De Broglie--Bohm:}]{0}
\bcbar{0}\smallskip

\bcbar[value=,text={\emph{d.} Everett (many worlds and/or many minds):}]{0}
\bcbar{18}\smallskip

\bcbar[value=,text={\emph{e.} Information-based/information-theoretical:}]{0}
\bcbar{24}\smallskip

\bcbar[value=,text={\emph{f.} Modal interpretation:}]{0}
\bcbar{0}\smallskip

\bcbar[value=,text={\emph{g.} Objective collapse (e.g., GRW, Penrose):}]{0}
\bcbar{9}\smallskip

\bcbar[value=,text={\emph{h.} Quantum Bayesianism:}]{0}
\bcbar{6}\smallskip

\bcbar[value=,text={\emph{i.} Relational quantum mechanics:}]{0}
\bcbar{6}\smallskip

\bcbar[value=,text={\emph{j.} Statistical (ensemble) interpretation:}]{0}
\bcbar{0}\smallskip

\bcbar[value=,text={\emph{k.} Transactional interpretation:}]{0}
\bcbar{0}\smallskip

\bcbar[value=,text={\emph{l.} Other:}]{0}
\bcbar{12}\smallskip

\bcbar[value=,text={\emph{m.} I have no preferred interpretation}]{0}
\bcbar{12}

\bcxlabel{percent of votes}
\end{bchart}}

\noindent The Copenhagen interpretation still reigns supreme here, especially if we lump it together with intellectual offsprings such as information-based interpretations and the Quantum Bayesian interpretation. In Tegmark's poll \cite{Tegmark:1998:qq}, the Everett interpretation received 17\% of the vote, which is similar to the number of votes (18\%) in our poll. Of course, given the different compositions of the respondent samples in the two polls, the numerical agreement with Tegmark's result might be merely coincidental. Similarly, the fact that de~Broglie--Bohm interpretation did not receive any votes may simply be an artifact of the particular set of participants we polled. Finally, looking back, we regret not to have included the ``shut up and calculate'' interpretation \cite{Mermin:1989:tk} (see also \cite{Fuchs:2000:az}) in our poll.

\subsubsection*{Question 13: How often have you switched to a different interpretation?}

\scalebox{0.85}{\begin{bchart}[step=10,max=100,unit=\%,width=0.8\textwidth]

\bcbar[value=,text={\emph{a.} Never:}]{0}
\bcbar{33}\smallskip

\bcbar[value=,text={\emph{b.} Once:}]{0}
\bcbar{21}\smallskip

\bcbar[value=,text={\emph{c.} Several times:}]{0}
\bcbar{21}\smallskip

\bcbar[value=,text={\emph{d.} I have no preferred interpretation:}]{0}
\bcbar{21}

\bcxlabel{percent of votes}
\end{bchart}}

\noindent Remarkably, one in three respondents professes to have \emph{never} switched interpretation. At the other end of the spectrum, one  respondent reported, by write-in, that he sometimes switches interpretation several times per day.

\subsubsection*{Question 14: How much is the choice of interpretation a matter of personal philosophical prejudice?}

\scalebox{0.85}{\begin{bchart}[step=10,max=100,unit=\%,width=0.8\textwidth]

\bcbar[value=,text={\emph{a.} A lot:}]{0}
\bcbar{58}\smallskip

\bcbar[value=,text={\emph{b.} A little:}]{0}
\bcbar{27}\smallskip

\bcbar[value=,text={\emph{c.} Not at all:}]{0}
\bcbar{15}

\bcxlabel{percent of votes}
\end{bchart}}

\noindent Debates about the foundations of quantum mechanics are sometimes perceived as an ideological battle where the objectivity of the scientific enterprise succumbs to philosophical and personal preferences. In our poll, a clear majority sees at least some influence of philosophical prejudices on the choice of interpretation of quantum mechanics. Whether we should be pleased with this realization and the situation in quantum foundations it reflects is difficult to say. In the absence of empirical differences between the interpretations, it is only natural to conclude that one's decision which interpretation to adopt will be influenced by personal preferences and beliefs. Of course, some may hope that this situation will be merely temporary, and that we should strive to settle the question of interpretation in a definitive way. For more on the topic, see Chapter~13 of \cite{Schlosshauer:2011:ee}.

\subsubsection*{Question 15: Superpositions of macroscopically distinct states}

\scalebox{0.85}{\begin{bchart}[step=10,max=100,unit=\%,width=0.8\textwidth]

\bcbar[value=,text={\emph{a.} Are in principle possible:}]{0}
\bcbar{67}\smallskip

\bcbar[value=,text={\emph{b.} Will eventually be realized experimentally:}]{0}
\bcbar{36}\smallskip

\bcbar[value=,text={\emph{c.} Are in principle impossible:}]{0}
\bcbar{12}\smallskip

\bcbar[value=,text={\emph{d.} Are impossible due to a collapse theory:}]{0}
\bcbar{6}

\bcxlabel{percent of votes}
\end{bchart}}

\noindent Our result reflects remarkable intellectual progress. Not long ago, the macroscopic--microscopic border was routinely conflated with the classical--quantum border, even though it had already been pointed out by Bohr that they should be regarded as independent concepts \cite{Schlosshauer:2008:im}. It may be that the continuing experimental progress in demonstrating quantum interference effects for ever-larger objects has changed the attitudes of many physicists.

\subsubsection*{Question 16: In 50 years, will we still have conferences devoted to quantum foundations?}

\scalebox{0.85}{\begin{bchart}[step=10,max=100,unit=\%,width=0.8\textwidth]

\bcbar[value=,text={\emph{a.} Probably yes:}]{0}
\bcbar{48}\smallskip

\bcbar[value=,text={\emph{b.} Probably no:}]{0}
\bcbar{15}\smallskip

\bcbar[value=,text={\emph{c.} Who knows:}]{0}
\bcbar{24}\smallskip

\bcbar[value=,text={\emph{d.} I'll organize one no matter what:}]{0}
\bcbar{12}

\bcxlabel{percent of votes}
\end{bchart}}

\noindent Should those who answered ``probably yes'' be proven right, then it would be fascinating to conduct another such poll 50 years from now. Notable write-ins included ``I won't be here,'' and ``I hope not.''

\section{\label{sec:corr}Correlations}

We also looked for relationships (correlations) between answers to different questions. To ensure representative sample sizes, we required a specific answer $A$ to have been checked by at least 4 participants. Then, if a fraction $f$ of members of this group had also checked a certain answer $B$, we registered a relationship between the two answers $A$ and $B$ if the following conditions were met:

\begin{enumerate}

\item[(i)] Answer $B$ is different from answer $A$ and was checked by at least 4 participants;

\item[(ii)] $f$ exceeded a (fixed) \emph{threshold} value $T$; 

\item[(iii)] $f$ exceeded the average percentage vote for answer $B$ among all 33 participants by a margin given by a (fixed) \emph{gap} parameter $G$. 

\end{enumerate}

The reason for imposing the last condition is that certain answers received a large number of votes across the board. For example, a total of 76\% of respondents agreed with the statement that ``quantum information is a breath of fresh air for quantum foundations.'' Condition (iii) therefore ensures that a relationship is only detected if the number of votes for $B$ cast by the members of the group who have voted for $A$ is in fact significantly different from the average of votes for $B$.
 
The larger $T$ and $G$ are chosen, the more significant (stronger) the correlations will be. We determined correlations for three different choices of parameter pairs $(T,G)$: $T=100\%$, $G=20\%$ (strong correlations); $T=75\%$, $G=15\%$ (medium correlations); and $T=50\%$, $G=10\%$ (weak correlations). 
 
\begin{figure}
\begin{center}
\includegraphics[width=\textwidth]{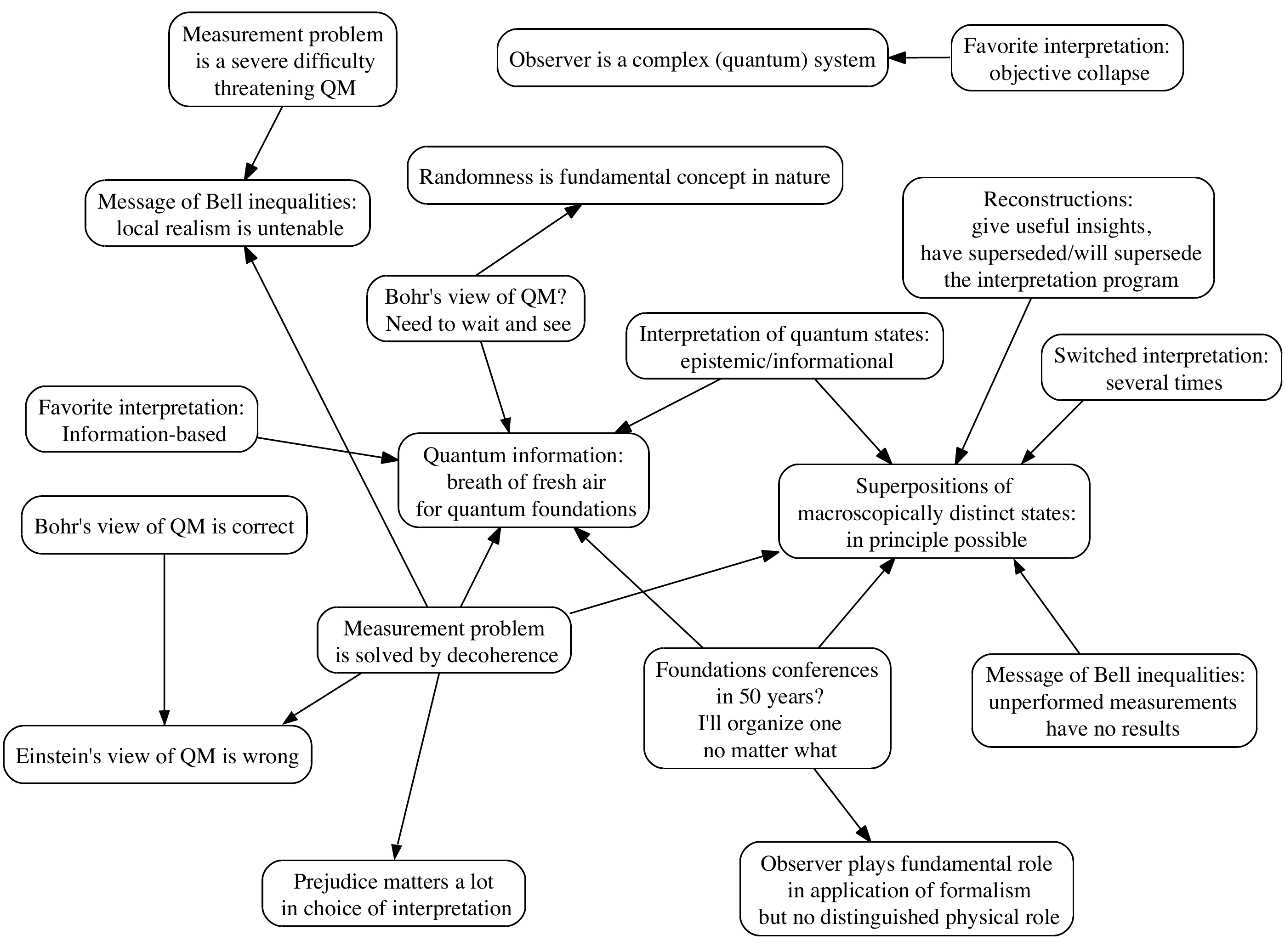}
\end{center}
\caption{\label{fig:xtrastrong}Strong correlations between answers. A relationship $A \rightarrow B$ between two answers $A$ and $B$ means that every respondent who checked $A$ also checked $B$, and that $B$ received less than 80\% of the total vote among all 33 participants of the poll.}
\end{figure}

\begin{figure}
\begin{center}
\includegraphics[width=\textwidth]{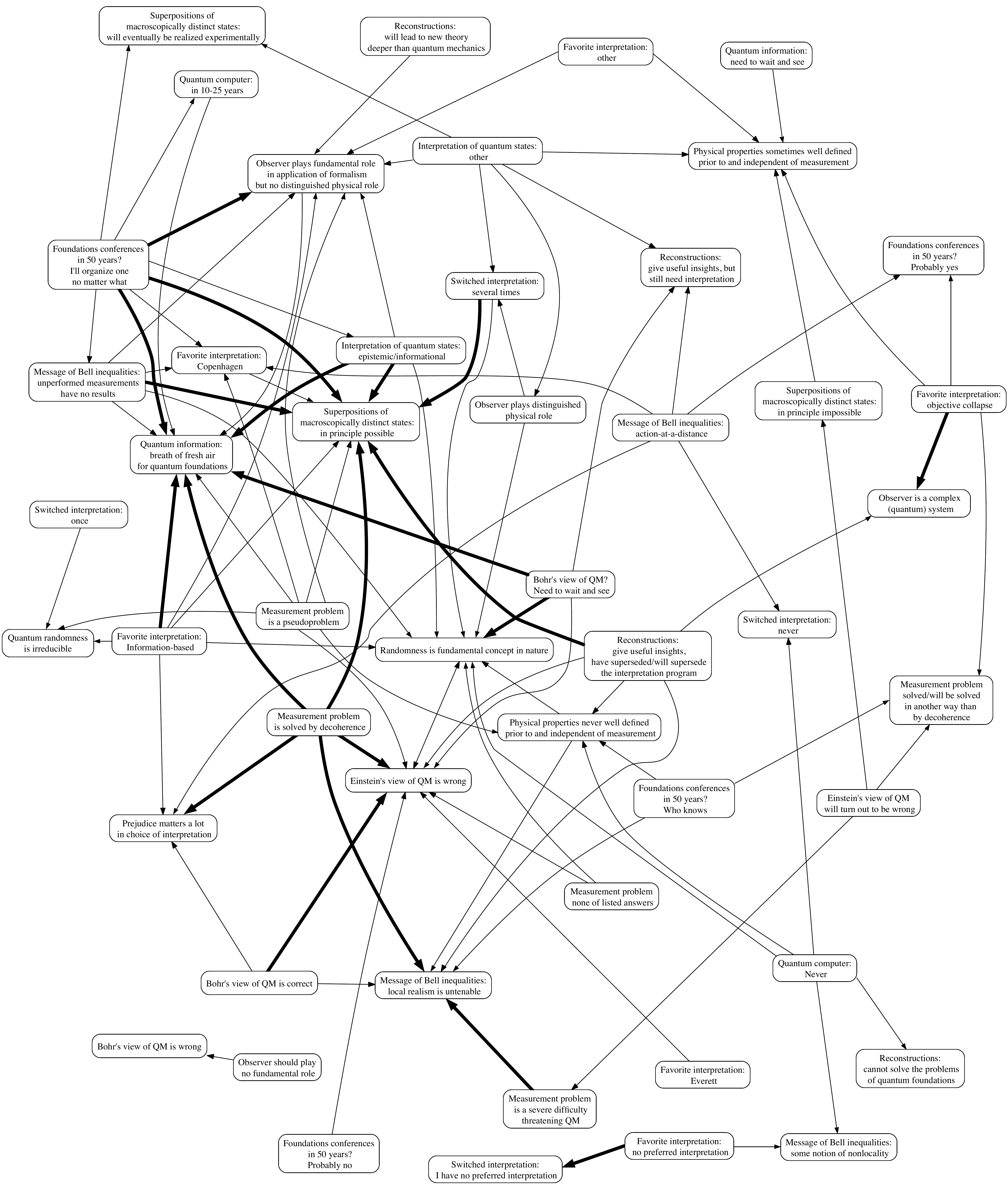}
\end{center}
\caption{\label{fig:mediumcorrels}Strong and medium correlations between different answers. Strong correlations are indicated by bold arrows, and are also separately visualized in Fig.~\ref{fig:xtrastrong}.}
\end{figure}
 
The strong correlations are shown in Fig.~\ref{fig:xtrastrong}; the medium correlations, which also include the strong correlations, are shown in Fig.~\ref{fig:mediumcorrels}. Due to the large number of weak correlations, a graph-like visualization proved not feasible in this case. Instead, Appendix~\ref{sec:correlation-table} presents these weak correlations (together with the strong and medium correlations) in form of a correlation matrix. 

Note that our definition of a relationship is not symmetric. Answer $A$ may be connected with answer $B$, but $B$ might not be connected with $A$, and vice versa. While we found many symmetric weak correlations---i.e., weak relationships between two answers matching conditions (i)--(iii) in both directions---there was only one symmetric medium correlation, and no symmetric strong correlation. The symmetric medium correlation was between the statements ``Randomness is a fundamental concept in nature'' and ``Einstein's view of quantum mechanics is wrong.'' While this particular correlation is not surprising, it \emph{is} perhaps surprising that there was not a larger number of symmetric medium correlations.

\section{\label{sec:discussion}Discussion}

The statements that found the support of a majority---i.e., answers checked by more than half of the participants---were, in order of the number of votes received:

\bn

\item Quantum information is a breath of fresh air for quantum foundations (76\%).

\item Superpositions of macroscopically distinct states are in principle possible (67\%). 

\item Randomness is a fundamental concept in nature (64\%).

\item Einstein's view of quantum theory is wrong (64\%).

\item The message of the observed violations of Bell's inequalities is that local realism is untenable (64\%).

\item Personal philosophical prejudice plays a large role in the choice of interpretation (58\%).

\item The observer plays a fundamental role in the application of the formalism but plays no distinguished physical role (55\%).

\item Physical objects have their properties well defined prior to and independent of measurement in \emph{some} cases (52\%).

\item The message of the observed violations of Bell's inequalities is that unperformed measurements have no results (52\%).

\en

Our results show a great diversity of opinion, transcending the traditional lines of Bohr-versus-Einstein, nonrealist-versus-realist, and epistemic-versus-ontic. To be sure, many correlations between answers still evidence traditional partisanship, but the field also seems infused with a new dynamism. Quantum information is hailed by a large majority as a ``breath of fresh air for quantum foundations,'' with information-based interpretations ranking second in the poll and epistemic-informational views of the quantum state beating strictly ontic interpretations. A majority of respondents found reconstruction approaches to quantum theory useful. Only one in four respondents sees the measurement problem as a ``severe difficulty threatening quantum mechanics,'' certainly an indication of shifting attitudes.

Many of the correlations we found (see Figs.~\ref{fig:xtrastrong} and \ref{fig:mediumcorrels}, and Appendix~\ref{sec:correlation-table}) represent familiar patterns, while others are more subtle and perhaps less expected. Let us mention a few examples of significant (medium or strong) correlations.

\begin{itemize}

\item The belief that superpositions of macroscopically distinct states are in principle possible found \emph{unanimous} support among those who preferred an epistemic/informational interpretation of quantum states; those who found reconstruction approaches not only useful but also thought that they had superseded, or will supersede, the interpretation program; those who declared to have switched interpretation several times;  those who regarded the measurement problem as solved by decoherence; and those who saw the message of Bell's inequalities as supporting Asher Peres's slogan that ``unperformed experiments have no results'' \cite{Peres:1978:aa}.

\item Those who believed that observers should not play any fundamental role whatsoever also tended to oppose Bohr's view of quantum mechanics. 

\item Those who dismissed the measurement problem as a pseudoproblem also tended to: embrace the Copenhagen interpretation; regard quantum randomness as fundamental and irreducible; reject Einstein's view of quantum mechanics; consider superpositions of macroscopically distinct states possible; and attribute a distinct formal (but not physical) role to the observer. Many of these attitudes were shared by those who preferred an information-based interpretation of quantum states and quantum theory. 

\item The idea that physical properties are ``sometimes'' well defined prior to and independent of measurement found broad support among those who preferred a physical-collapse theory; those who regarded superpositions of macroscopically distinct states as in principle impossible; and  those who thought that we will ``have to wait and see'' when it comes to judging the relevance of quantum information for the foundations of quantum mechanics. These correlations may point to a realist attitude toward quantum mechanics.

\item All of those who regarded the measurement problem as solved by decoherence also embraced quantum information, thought that superpositions of macroscopically distinct states are possible, rejected Einstein's view of quantum mechanics wrong, considered local realism as untenable, and believed that philosophical prejudice plays a large role in the choice of interpretation.

\item Many of those who categorically answered ``Never!''\ to the question of when we may have a working and useful quantum computer were also skeptical of reconstructions of quantum theory (``Reconstructions cannot solve the problems of quantum foundations'') and stated to have never switched to a different interpretation. 

\item Among the different interpretive camps, adherents of objective (physical) collapse theories were the only group to believe, in significant numbers, that in fifty years from now, there will likely be still conferences devoted to quantum foundations. So perhaps this reflects the fact that those who pursue collapse theories tend to view quantum theory as an essentially unsatisfactory and unfinished edifice requiring long-term modification and construction efforts. Vice versa, it may be a sign that those who regard such efforts as unnecessary or even misguided are optimistic that the remaining foundational problems, whatever they may be, will soon be resolved. 

\end{itemize}

\section{Conclusions}

Quantum theory is based on a clear mathematical apparatus, has enormous significance for the natural sciences, enjoys phenomenal predictive success, and plays a critical role in modern technological developments. Yet, nearly 90 years after the theory's development, there is still no consensus in the scientific community regarding the interpretation of the theory's foundational building blocks. Our poll is an urgent reminder of this peculiar situation. 

\subsubsection*{Acknowledgements}
We thank all participants for their willingness to share their views. The conference where the poll was conducted was made possible through generous financial support by the John Templeton Foundation.

\newpage

\appendix 

\section{\label{sec:correlation-table}Correlation table}

The table below lists correlations between different answers, using the criteria and parameters described in Sec.~\ref{sec:corr}. Strong correlations are shown in red and additionally labeled by the number ``3''; medium correlations are shown in orange and labeled by ``2''; weak correlations are shown in yellow and labeled by ``1.'' The first column (labeled by ``\#'') displays the number of votes received for the answer listed in the second column (labeled by $A$). The answer labels refer to the bar graphs of Sec.~\ref{sec:results}; for example, ``1c'' refers to answer \emph{c} (``The randomness is irreducible'') of Question 1 (``What is your opinion about the randomness of individual quantum events?''). The two top rows display the question titles and the target answers (labeled by $B$). For every answer $A$, the corresponding row shows its correlations with the different answers $B$. For example, the first row shows that answer \emph{c} of Question 1 was checked 16 times and is weakly correlated (threshold $T=50\%$, gap $G=10\%$) with answers 2b, 5a, 12b, and 14a.

\begin{center}
\includegraphics[width=\textwidth]{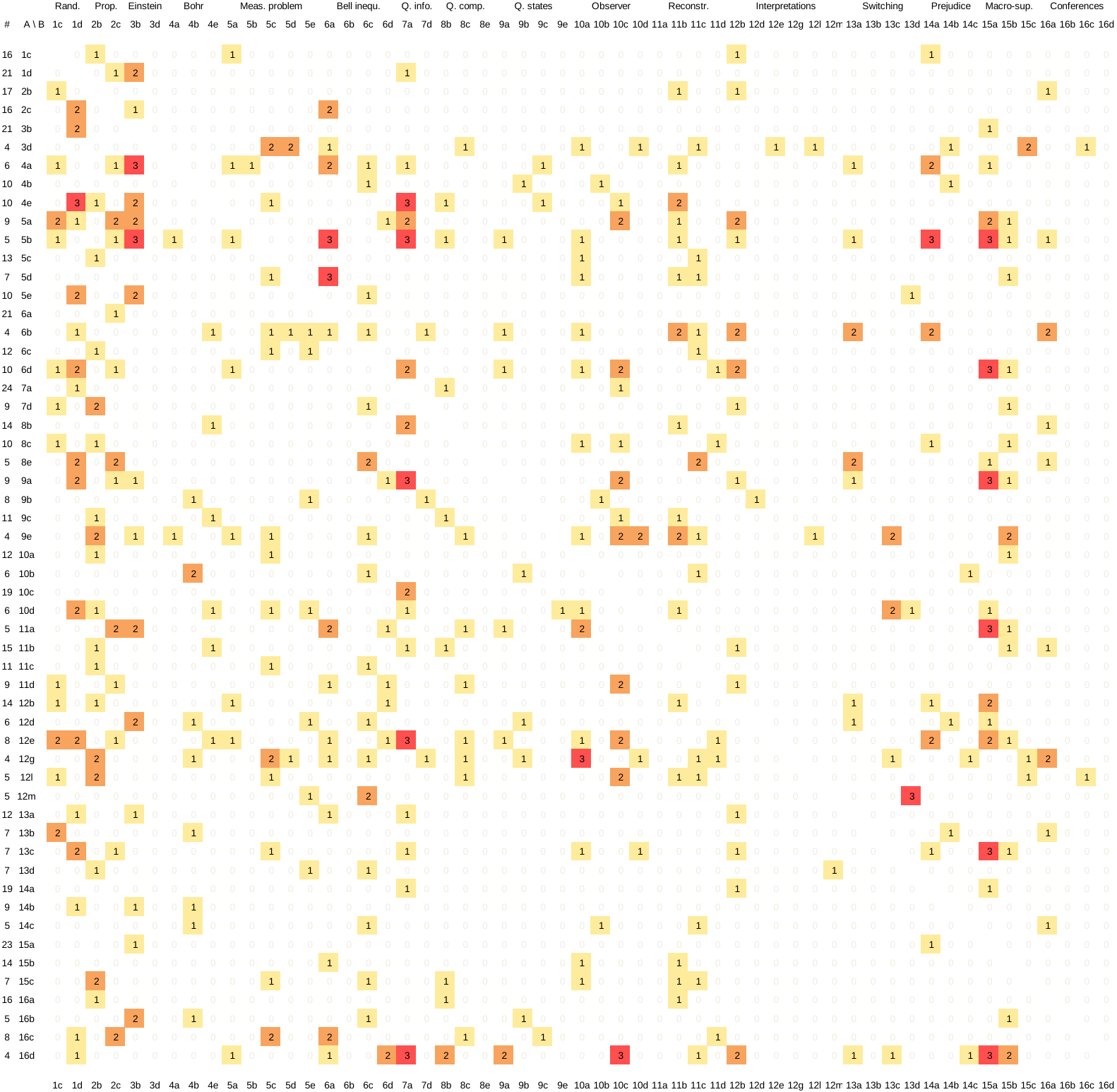}
\end{center}

\section{\label{sec:list-conf-part}List of conference participants}

Below we list the names and affiliations of the participants of the conference ``Quantum Physics and the Nature of Reality,'' held July 3--7, 2011, at the International Academy Traunkirchen, Austria. Of these 35 participants, 33 completed our questionnaire.\\[-.1cm]

Markus Aspelmeyer (University of Vienna)

Charles Bennett (IBM Research Division)

Tina Bilban (University of Vienna \& Austrian Academy of Sciences)

Andrew Briggs (University of Oxford)

{\v{C}}aslav Brukner (University of Vienna)

Jeffrey Bub (University of Maryland)

Jeremy Butterfield (University of Cambridge)

Thomas Durt (Griffith University)

George Ellis (University of Cape Town)

Robert Fickler (University of Vienna \& Austrian Academy of Sciences)

Christopher Fuchs (Perimeter Institute for Theoretical Physics)

Rodolfo Gambini (Louisiana State University)

Daniel Greenberger (City College of New York)

Richard Healey (University of Arizona)

Adrian Kent (University of Cambridge) 

Simon Kochen (Princeton University)

Johannes Kofler (University of Vienna \& Austrian Academy of Sciences)

Paul Kwiat (University of Illinois at Urbana--Champaign)

Fotini Markopoulou (Perimeter Institute for Theoretical Physics)

Argyris Nicolaidis (Aristotle University of Thessaloniki)

Bill Plick (University of Vienna \& Austrian Academy of Sciences)

Robert Polster (University of Vienna \& Austrian Academy of Sciences)

Sandu Popescu (University of Bristol)

Sven Ramelow (University of Vienna \& Austrian Academy of Sciences)

Jess Riedel (Santa Fe Institute)

Simon Saunders (University of Oxford)

Tejinder Singh (Tata Institute of Fundamental Research)

Andrew Steane (University of Oxford)
 
Stig Stenholm (KTH Royal Institute of Technology)

Jos Uffink (University of Utrecht) 

Umesh Vazirani (University of California, Berkeley)

Vlatko Vedral (University of Oxford)

Alexandre Zagoskin (Loughborough University)

Anton Zeilinger (University of Vienna \& Austrian Academy of Sciences)

Marek {\.{Z}}ukowski (University of Gdansk)

\end{document}